\documentclass[12pt,preprint]{aastex}

\newcommand{\feh} {\mbox{\rm [Fe/H]}}
\newcommand{\teff}  {\mbox{$T_{\rm eff}$}}
\newcommand{\logg}  {\mbox{{\rm log} $g$}}
\newcommand{\Msun}  {\mbox{$M_\sun$}}
\newcommand{\kmprs}  {\mbox{\rm\,km\,s$^{-1}$}}

\shorttitle{Li abundances between stars with and without planets}
\shortauthors{Chen and Zhao}

\begin{document}

\title{A Comparative Study on Lithium Abundances in Solar--Type Stars With and Without Planets}

\author{Y. Q. Chen and G. Zhao} 
\affil{National Astronomical Observatories, Chinese Academy of Sciences,
A20, Datun Road, Chaoyang District, Beijing 100012, China; gzhao@bao.ac.cn}

\begin{abstract}
We have investigated the abundance anomalies of lithium for stars with planets
in the temperature range of 5600--5900\,K reported by Israelian and coworkers, 
as compared to 20 normal stars at the same temperature and metallicity
ranges.
Our result indicates a higher probability of
lithium depletion for stars with planets in the main--sequence stage.
It seems that stellar photospheric abundances of lithium
in stars with planets may be somewhat affected by the presence of
planets. Two possible mechanisms are considered to account
for the lower Li abundances of stars with planets. One is
related to the rotation-induced
mixing due to the conservation
of angular momentum by the protoplanetary disk, and the
other is a shear
instability triggered by planet migration.
These results provide new information on stellar evolution
and the lithium evolution of the Galaxy.
\end{abstract}


\section{Introduction}
Since the first announcement of a planet orbiting a Sun-like
star, 51 Peg, by \citet{Mayor95}, around 140 stars with planets
(SWPs) have been reported. The main properties of SWPS
have been investigated by many works. Among these,
spectroscopic observations of most SWPs based on high-resolution 
and high signal-to-noise ratio spectra have been carried out. 
All of these works indicate a high mean
metallicity of SWPs relative to normal field stars. Are there
any other chemical signatures for SWPs? There is no conclusive
result on this issue, and only a few works have reported abundance
anomalies in some elements. For example, \citet{Gonzalez00}
suggested a high [C/Fe] ratio for SWPs as compared with the normal stars from \citet{Gustafsson99}.
However, they later reported that there is no
difference in abundance for C, Na, Mg, and Al elements between SWPs and normal stars \citep{Gonzalez01}. Present
knowledge on the abundances of 25 elements (except for Li and  Be) available in the
literature shows no special feature for SWPs except for the high overall metallicity.

For lithium abundance, \citet{Israelian01}
claimed that $^6{\mathrm Li}$ was detected in the atmosphere of HD
82943, while \citet{Reddy02} did not confirm this detection.
\citet{Mandell04} reported no $^6 \mathrm {Li}$ presence in several
lithium-poor stars. For $^7\mathrm {Li}$, \citet{King97} first found several SWPs
with low Li abundances, and they related them with the presence of a planetary companion.
 \citet{Gonzalez00} concluded that
smaller Li abundances are found for SWPs after correcting
for trends with temperature, metallicity, and age.
However, \citet{Ryan00} suggested that there is no significant
difference in lithium abundance for SWPs, and later work by
\citet{Gonzalez01} agreed with Ryan's conclusion. With a large
sample, \citet{Israelian04} suggested again that in the
temperature range
5600--5850\,K, SWPs show additional depletion of Li
compared to normal stars. They also showed that for higher temperatures,
the Li abundances were similar for SWPs and for normal
stars.
Recently, measurements
of Be abundances by \citet{Santos04} indicated further evidence
on the composition difference between stars with and without planets.

Lithium is a special element because it is easily destroyed in
stars during even pre--main sequence and main sequence evolution, which
is not the case for other elements. Therefore, Li abundance reflects
the mixing history of the star. For SWPs, Li is
further related with the accretion of material and the
angular--momentum evolution of the system, and thus
Li abundances have been used to distinguish the current hypotheses on
planet formation mechanisms. In this respect, it is
important to verify whether Li abundance anomalies for SWPs at the temperature range of
5600--5850\,K reported by \citet{Israelian04} are real,
which will provide new information on the planet formation
process and its influence on host stars.

In this work we conduct a parallel abundance analysis of lithium
abundances for 16 SWPs and 20 comparison stars in the same
metallicity of $\feh \sim 0.1$ dex at the temperature range of
5600--5900\,K in order to investigate whether any anomalies persist.
Furthermore, this study is quite interesting for our
understanding of the lithium behavior of solar-- metallicity stars,
which show a large scatter at $5600 < \teff < 5900$ K and which the
standard model of stellar evolution cannot explain.
With a consistent analysis procedure, this work aims
to investigate the effect of planet presence on stellar evolution.
This will provide new information on the chemical evolution of lithium
in the Galaxy.

\section{Observations and Abundance Analysis}

SWPs are selected from a catalog \footnote{http://www.obspm.fr} in
the temperature range of 5600--5900\,K, and their spectra
are available from our observations with the
2.16m telescope of the National Astronomical Observatories (Xinglong, China)
with a resolving power of 37\,000, or from \citet{Prugniel01}, who
observed with the 1.93\,m telescope of the Observatoire de Haute Provence
with R$\sim$42\,000. \citet{Prugniel01} present a database of
high--resolution spectra of 709 stars covering a large range of
atmospheric parameters, which enables us to select a comparison
sample of stars spanning the same ranges in temperature, gravity
and metallicity as the above sample of planet stars but not having been
reported to harbor planets. That is,
both samples cover the atmospheric parameters of
$5600 < \teff < 5900$ K, $\logg > 3.8$,
and $\feh > -0.3$.
In the selection, str\"omgren $uvby$ indices from \citet{Olsen83,Olsen93} are adopted and the metallicity is estimated based on
the calibrations by \citet{Schuster89}. Along with the initial metallicity,
temperature and gravity are derived in the same way as
described below.
All spectra have signal-to-noise ratios above 150 pixel$^{-1}$
at the \ion{Li}{1} $\lambda$6707 line.

The spectra were reduced using standard MIDAS (for Xinglong data)
and IRAF (for ELODIE data) routines for order definition,
background correction, flat-fielding, extraction of echelle
orders, wavelength calibration, and continuum fitting. The
\ion{Fe}{2} lines suitable for measurement were carefully selected,
and the equivalent widths (EWs) were obtained by the fitting of a
Gaussian function. Atomic data for \ion{Fe}{2}
lines were the same as those in
\citet{Chen02}. The wavelengths and oscillator strengths of the
\ion{Li}{1} $\lambda$6707 line are taken from \citet{Smith98}.

\begin{table}[]
\caption[]{Stellar Parameters for SWPs and the Comparison Stars}
\label{Tab:Liabu}
\begin{tabular}{crccrrrrrrrr}
\hline\noalign{\smallskip}
Star &  $\teff$   & $\logg$ & $\xi_t$ & $\feh$ & Li & Age & EW(1) & $\sigma_W$   &EW(2) &EW(3) &EW(4) \\
HD   &  K         &         & $\kmprs$ &       &     & Gyr & mA & mA & mA & mA &mA \\
\hline\noalign{\smallskip}
\multicolumn{12}{c}{Stars with planets}\\
  Sun   &  5780&  4.44&  1.15 &   $-$0.03&$< 1.27$ & 4.6&...   & 1.2 & ... & ...& ...\\
  12661 &  5682&  4.26&  1.50 &    0.30&   1.74  & 9.6&16.6 & 0.8& ... & ...& ...\\
  16141 &  5752&  4.13&  1.40 &    0.08& $<1.03$ & 7.6& ...  & 1.9& $<2.0$ & ... & ...\\
  23596 &  5906&  4.09&  1.60 &    0.31&   2.78  & 5.9&86.7 & 1.6& ... & ...& ...\\
  33636 &  5817&  4.41&  1.30 &   $-$0.03&   2.29  & 5.8&40.3 & 1.5& 49.0 & ...& ...\\
  72659 &  5846&  4.16&  1.15 &    0.04&   2.25  & 6.3&35.4 & 0.6& ... & ...& ...\\
  82943 &  5858&  4.32&  1.40 &    0.20&   2.37  & 6.0&44.7 & 2.2&44.1 & ...& ...\\
  92788 &  5683&  4.36&  1.40 &    0.18&$<0.97$  & 6.6&...   & 1.7&... & ...& ...\\
  95128 &  5755&  4.23&  1.30 &   $-$0.00&   1.68  & 9.3& 10.5& 2.0&12.5& ...&18.0 \\
 106252 &  5769&  4.32&  1.00 &    0.01&   1.72  & 9.6&13.5 & 2.1& ... & ...& ... \\
 134987 &  5735&  4.27&  1.15 &    0.24& $<1.02$ & 8.3&... & 1.2&$<2.0$ & ...& ...\\
 143761 &  5701&  4.26&  1.20 &   $-$0.21&   1.37  &10.8&6.2 & 1.1& 5.9 & ...& 6.0\\
 150706 &  5764&  4.43&  1.20 &    0.01&2.40  &  7.5& 53.3&  1.4& 55.4& ...& ...\\
 187123 &  5717&  4.26&  1.15 &    0.11&$<1.00 $ & 8.8& ...  & 1.8 &... & ...& ...\\
 195019 &  5729&  4.07&  1.15 &    0.03&   1.55  & 7.3& 9.9 & 1.8& 6.9 & ...& ...\\
 217014 &  5654&  4.27&  1.15 &    0.20&$<0.89$ &10.4& ...  &  0.8&4.4 & ...& ...\\
\hline\noalign{\smallskip}
\multicolumn{12}{c}{The comparison stars}\\
   4307 &  5748&  3.95&  1.30 &   $-$0.20&   2.38 & 6.9&53.9 & 1.9& 52.3 & ...& ...\\
   4614 &  5818&  4.31&  1.20 &   $-$0.23&   2.08 &10.3&21.7 & 1.0& 26.3 & 24.0& 21.0\\
   9562 &  5786&  3.99&  1.20 &    0.18&  2.51  & 4.6&61.0 & 1.7&61.2 & ...& ...\\
  15335 &  5797&  3.93&  1.40 &   $-$0.15&   2.55 & 7.0&65.7 & 0.7& ... & 57.9& 52.0\\
  34411 &  5800&  4.24&  1.20 &    0.07&   2.07 & 7.2&24.3 & 1.1& 25.7 & 27.4& ...\\
  39587 &  5833&  4.41&  1.40 &   $-$0.00&   2.85 & 5.8&103.9& 1.1&101.3 &  105.6 &95.0\\
  52711 &  5778&  4.31&  1.30 &   $-$0.10&   1.84 & 9.2&17.2 & 1.5&17.5 & ...& ...\\
  70110 &  5880&  3.93&  1.40 &    0.10&   2.49 & 3.9&53.2 & 1.9&52.0 & 55.8& ...\\
  76151 &  5702&  4.39&  1.20 &    0.05&   1.80 & 6.0&18.2 & 1.6&18.7 & ...& ...\\
  79028 &  5818&  4.05&  1.50 &    0.03&   2.65 & 7.2&76.1 & 1.4&74.1 & 76.5& 11.0\\
  84737 &  5822&  4.11&  1.30 &    0.17&   2.32 & 5.7&42.9 & 1.5&41.8 & ...& ...\\
  88986 &  5750&  4.09&  1.30 &   $-$0.02&   1.97 & 7.2&23.6 & 1.5&26.2 & ...& ...\\
 109358 &  5742&  4.28&  1.20 &   $-$0.26&   1.59 &12.0&10.7 & 0.8& 9.3 & ...& 9.0\\
 114710 &  5920&  4.39&  1.30 &    0.13&   2.56 & 2.3& 63.3& 1.0&57.8 & ...& ...\\
 115383 &  5918&  4.19&  1.30 &    0.19&   2.78 & 5.2& 79.2& 1.3&83.1 & ...& ...\\
 141004 &  5806&  4.17&  1.40 &   $-$0.02&   1.79 & 8.2& 14.7& 1.5&16.0 & ...& 20.0\\
 182572 &  5648&  4.12&  1.40 &    0.15&$<0.90$ & 8.8& ...  & 2.4&$<1.0$ & ...& ...\\
 190406 &  5797&  4.38&  1.40 &    0.04&   2.26 & 5.6& 39.4& 1.8& 37.6 & ...& ...\\
 193664 &  5795&  4.39&  1.20 &   $-$0.11&   2.27 & 6.1& 39.9& 1.2&34.0 & ...& ...\\
 196755 &  5675&  3.66&  1.40 &   $-$0.04&   1.28 & 3.0& 6.6 & 1.3&10.8 & ...& ...\\
\hline\noalign{\smallskip}
\tablenotetext{ }{Stellar parameters, Li abundance, age, EWs, and  their errors
for SWPs and the comparison stars. The EWs of the Li line from \citet{Takeda05},
\citet{Chen01}, and \citet{Lambert91} are presented in columns 
EW (2), EW(3) and EW(4) respectively.}
\end{tabular}
\end{table}

\clearpage

\begin{figure}
\epsscale{1.0}         
\plotone{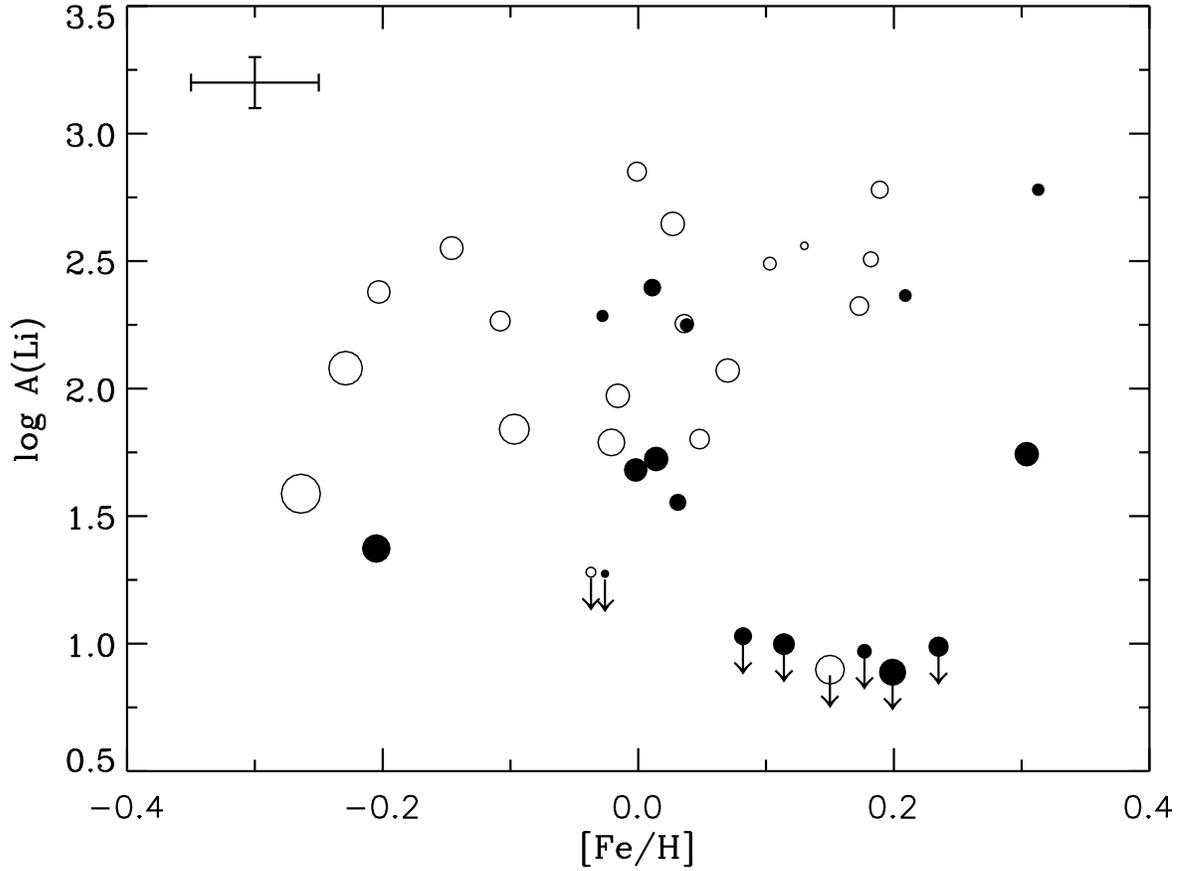} \figcaption{Li abundances vs.
[Fe/H] for stars with ({\it filled circles}) and without ({\it open circles}) planets.
Upper limits of Li abundances are indicated by downward--directed arrows,
and the size of the symbol corresponds to stellar age.} \label{Fig:plot1}
\end{figure}

\begin{figure}
\epsscale{1.0}         
\plotone{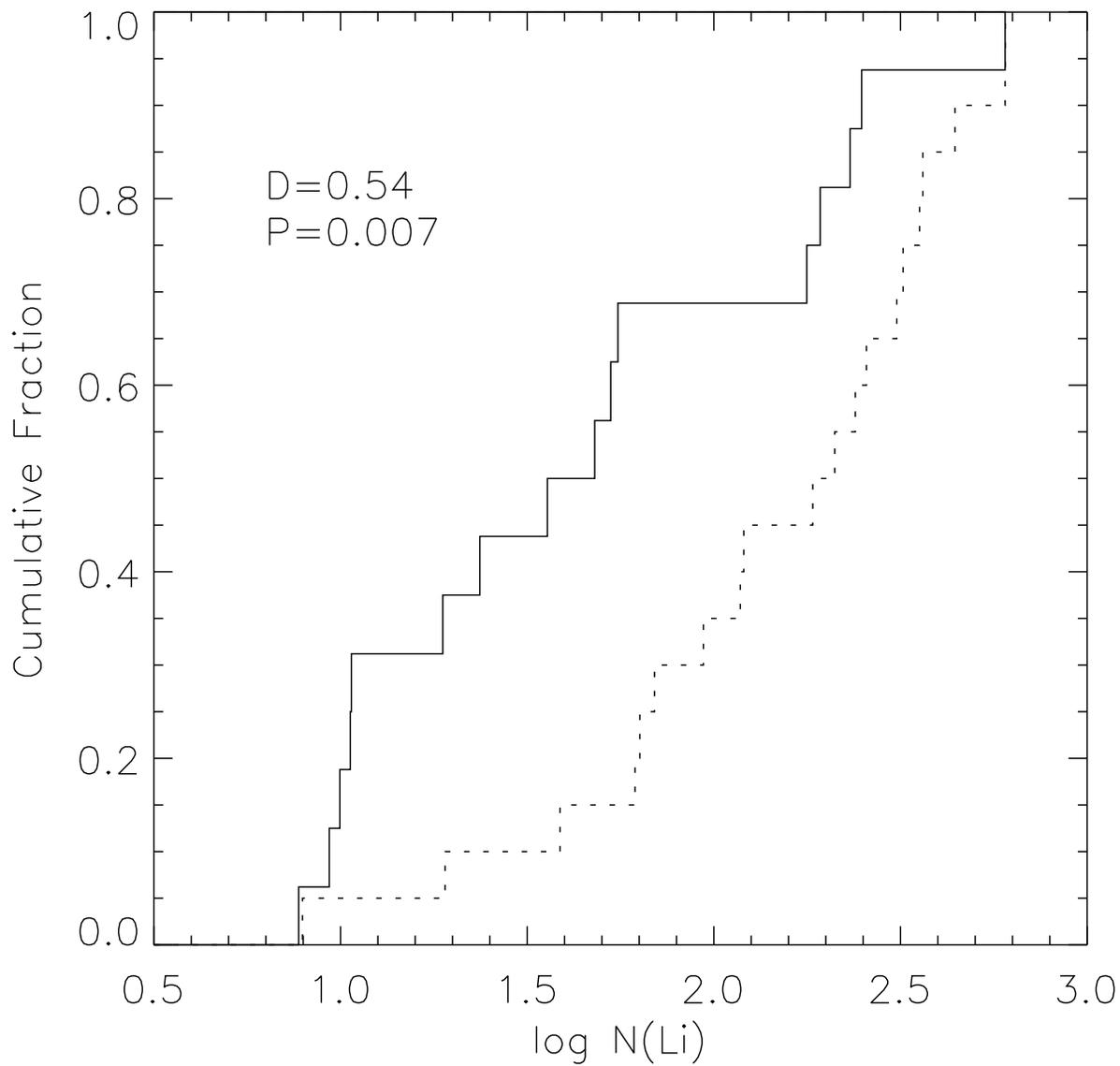} \figcaption{Cumulative distribution functions
based on the Kolmogorov-Smirnov test for stars with ({\it solid line})
and without ({\it dash line}) planets. Shown are different groups between the
two samples, with the maximum value of D = 0.54 and the
probability for the two samples being the same group of 0.007.}
\label{Fig:plot2}
\end{figure}

\begin{figure}
\epsscale{1.0}         
\plotone{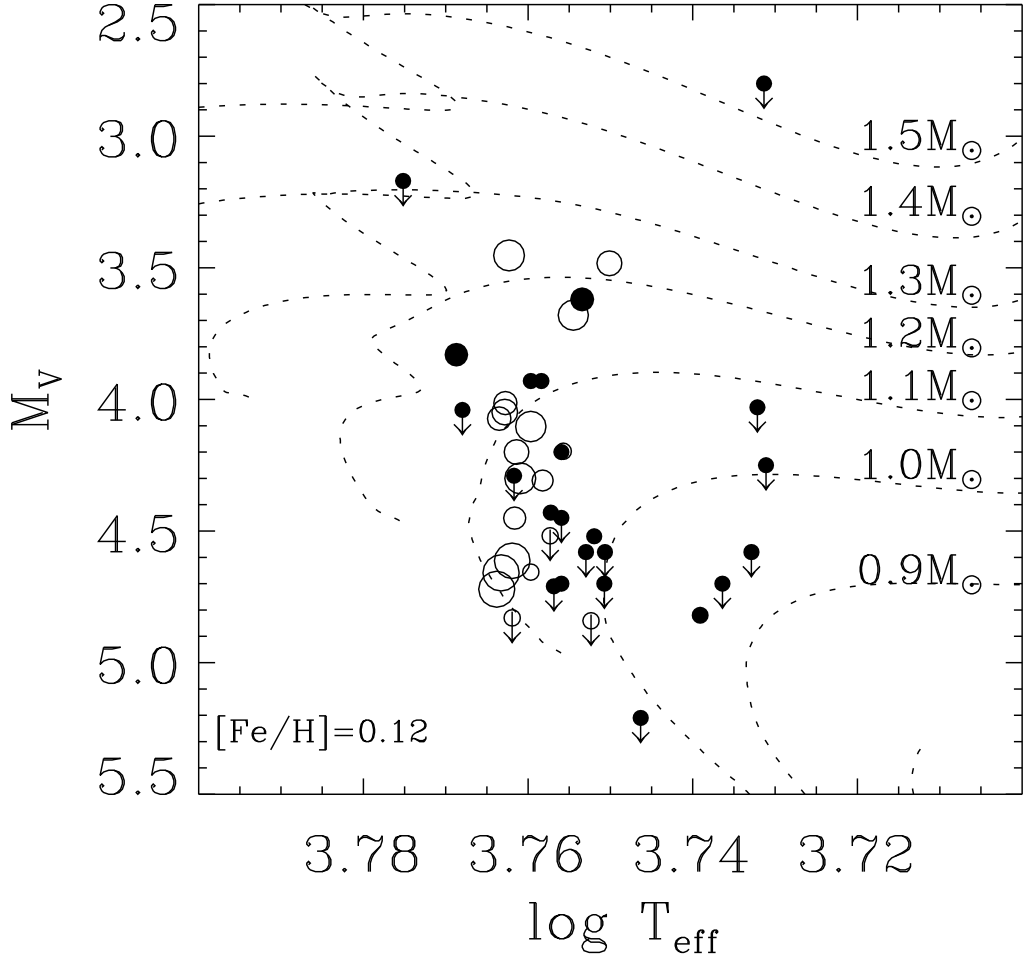}
\figcaption{Stellar positions of SWPs in \citet{Israelian04} ({\it filled circles}) with new determined temperatures as
compared with Chen et al. (2001; {\it open circles}) for stars with $5600 < \teff < 5850$ K and $\feh > -0.3$. Downward--directed arrows correspond to stars with upper limits for Li abundances, and diameters of circles correspond to stellar Li abundances taken directly from \citet{Israelian04} and \citet{Chen01}.}
\label{Fig:plot3}
\end{figure}

\begin{figure}
\epsscale{1.0}          
\plotone{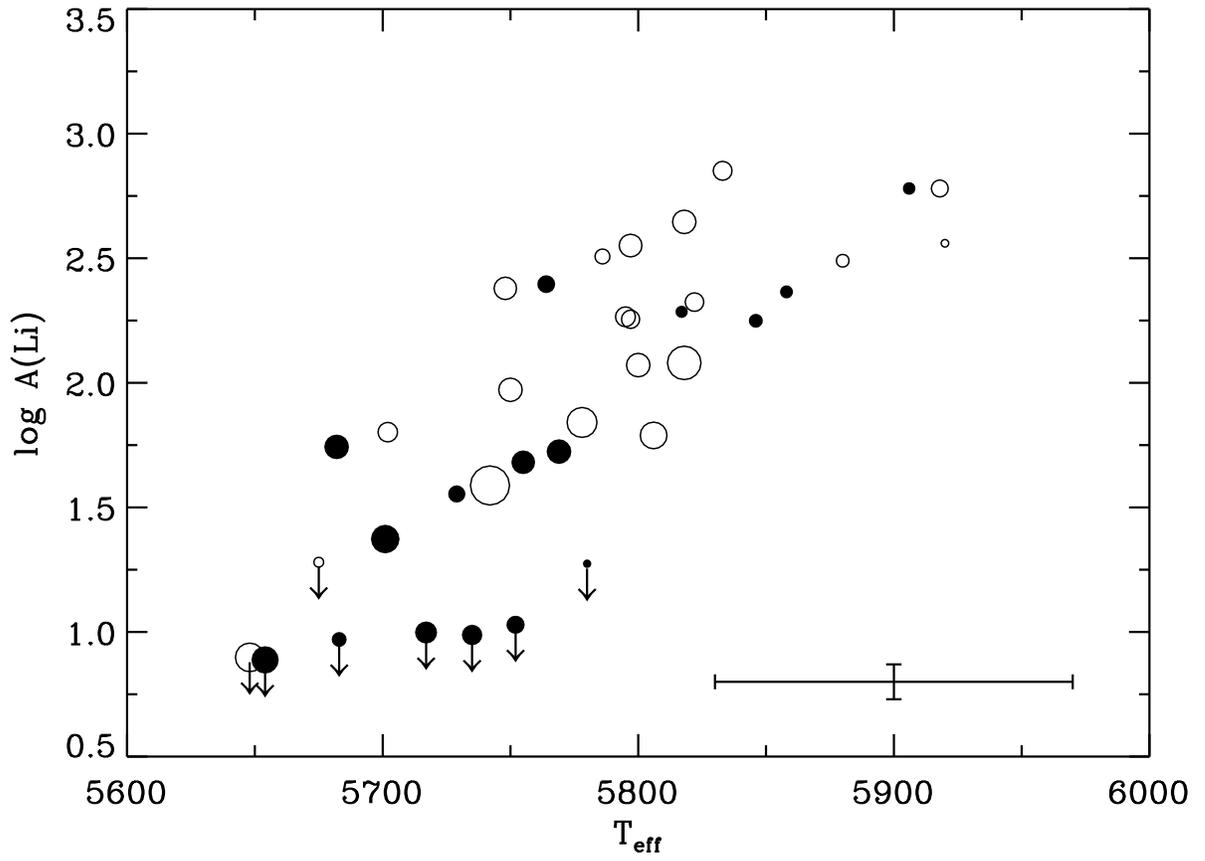} \figcaption{Li abundance vs.
$\teff$. The symbols are the same as in Fig.~1.}
\label{Fig:plot2}
\end{figure}

\begin{figure}
\epsscale{1.0}         
\plotone{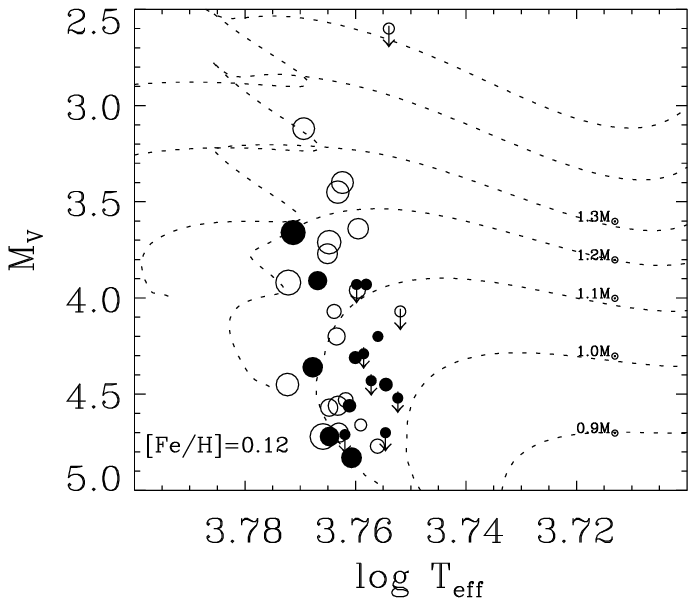}
\figcaption{Stellar positions in the HR diagram as compared with the evolution tracks from \citet{VandenBerg00}. The symbols are the same as in Fig.~1,
but diameters of circles correspond to Li abundances.}
\label{Fig:plot3}
\end{figure}

Effective temperatures were determined from the Str\"{o}mgren
$(b-y)$ color index \citep{Olsen83, Olsen93} using the calibration
of \citet{Alonso96}. The gravities were determined via {\it Hipparcos}
parallaxes \citep{Perryman97}, and the microturbulences were obtained
by forcing \ion{Fe}{1} lines with different strengths to give
the same abundances. {\bf Here the oscillator strengths of the \ion{Fe}{1} lines
are taken from \citet{Chen00}.}
The new metallicity derived from
spectroscopic analysis was updated, and the procedure of
atmospheric parameter determination was iterated for consistency,
even though there is no systematic difference between the
initial metallicity and the new value, with a deviation of
$0.01\pm0.15$ dex. The internal errors of the atmospheric parameters are
around 70\,K in temperature, 0.1 dex in $\logg$, 0.1 dex in $\feh$,
and 0.3 $\kmprs$ in microturbulence. But the absolute
uncertainties of the stellar parameters may be slightly larger.

We have found that the photometry-based temperatures in the work
are 130 K lower than the spectroscopically--derived values
in \citet{Santos04} for SWPs in common. However,
our temperatures are 83\, K higher than those in \citet{Takeda05}, who
derive temperatures by spectroscopic method for 29 stars
in common. It seems that
spectroscopically--derived temperatures have significant
uncertainties, and the values depend on stellar model atmosphere,
the selection of lines, the adopted atomic data, and the way the
temperatures are determined. In addition,
as already noted by \citet{Santos04}, surface gravities
determined from the balance of \ion{Fe}{1} and \ion{Fe}{2} abundances
are systematically higher than those derived from parallaxes (adopted in
our work) by the order of 0.15--0.20 dex. However, iron abundances
in this work and \citet{Santos04} are similar since we adopted
iron abundances from \ion{Fe}{2} lines and they forced
 \ion{Fe}{1} and \ion{Fe}{2} lines to give the same iron abundances
when they derived gravities. For the comparison stars, there are
15 stars in common with \citet{Edvardsson93}, and our temperatures
are 87\,K lower than their values. This is due to different temperature
calibrations, and our values, based on the infrared flux method temperatures by
\citet{Alonso96}, may be more reliable.
Actually, the absolute temperature
and gravity are not very important, since we are carrying out
a differential analysis, and it is crucial that the stellar parameters all be
in the same scale between the target and the comparison sample.

The model atmospheres were interpolated from a grid of
plane-parallel, LTE models provided by \citet{Kurucz95} in which
convective overshoot is switched off. The
ABONTEST8 program developed by P. Magain in the Li\'ege group
was used to carry out the calculations of theoretical line EWs, 
and abundances were derived by matching the
theoretical EWs to the observed values.
Abundance errors, estimated from the uncertainties
of atmospheric parameters and EWs, are
0.07 dex for iron abundances, and less than 0.1 dex for
Li abundances. Here the contribution of the $^6 \mathrm{Li}$ isotope
to the \ion{Li}{1} $\lambda$6707 line was assumed to be negligible.
Finally, non-LTE corrections
were applied to the derived Li abundances based on the work by
\citet{Carlsson94}, who studied non-LTE formation of the
\ion{Li}{1} $\lambda6708$ line as a function of effective
temperature, gravity, metallicity, and Li abundance. Stellar
parameters and Li and Fe abundances for SWPs and
the comparison stars, as well as the EWs
of the lines and their errors
are shown in Table 1.

\section{Results and Discussions}
\subsection{Li Abundances versus $\feh$}
Li abundances as a function of metallicity for all stars in the present
survey are shown in Fig. 1, in which
stars with downward--directed arrows have an undetected lithium line,
and the symbol size corresponds to stellar age.
The upper limits of the Li abundances are estimated by
assuming EWs of 3.0 m\AA, which is the estimated
error in EW from the comparison 
used in \citet{Chen00}. Based on Table 1,
the scatter of the EW comparison between this work
and \citet{Takeda05}, \citet{Chen01}, and \citet{Lambert91}
is around 4.0 m\AA, which indicates an error of $4.0/\sqrt{2} \sim 3.0$
m\AA\
assuming the same uncertainties for the EWs from the other works.
Note that the statistical uncertainties based on the formula
by \citet{Cayrel88}, presented in Table 1, are smaller.
Since experimental errors by different authors during spectrum 
normalization and EW measurements are not considered in
Cayrel's (1988) formula, we believe that
an upper limit to the EWs of 3.0 m\AA\ for
stars with an undetected Li line may be more reasonable.

An interesting result
from Fig.~1 is that there is a lack of normal stars with $\feh > +
0.2$. This suggestion is supported by large samples of stars
in the literature. There are 22\% normal
stars with $\feh > +0.2$ in \citet{Valenti05}'s sample at
the temperature range of 5600--5900\,K, but note that their
temperatures, derived from the fitting of synthetic spectra 
with observed spectra, are 107\,K higher than  our values 
for stars in common, and the sample is too small for statistics.
With photometry-based temperatures and a larger sample of stars
from \citet{Nordstrom04},
this fraction is reduced to 2\%, indicating a low probability
of super--metal--rich stars in the solar neighbourhood.
This is probably due to the fact that most metal--rich
stars tend to have formed planetary systems, and thus it is
difficult to obtain a comparison sample that consists of 
super--metal--rich
stars not harboring a planet. This is consistent with the current
knowledge that SWPs have higher mean metallicities than
normal stars.

The second feature from Fig.~1 is that six of the 16 SWPs
versus two of the 20 normal stars have undetected lithium
lines. All of the six SWPs with undetected Li lines are 
main--sequence stars while the two normal stars with depleted Li
abundances are subgiants, as shown below (Fig.~5).
It seems that main--sequence SWPs in the temperature
range of 5600--5900\,K destroy their Li much more easily
than do normal stars.
A Kolmogorov-Smirnov test is applied to stars with and without planets,
and the cumulative distribution functions are shown in Fig.~2.
It shows that the maximum value of the absolute difference
between the two distribution functions is 0.54, and the probability
for the two samples being the same group is 0.007.
This is also evident
from \citet{Israelian04}, as shown in Fig.~3, in which eight of 16 
main--sequence
SWPs at $5600 <\teff< 5850$ K only have upper limits for
Li abundances, while three of the 14 normal main--sequence stars from \citet{Chen01}
are depleted in Li. Note that the two comparison samples, from this work
and \citet{Chen01}, are independently selected with six common stars.
Thus, this property could be intrinsic, but further investigations
with large samples of both SWPs and normal stars are needed.

If this suggestion that main--sequence SWPs have a higher probability
of depleting their lithium abundances is true, it will provide new
information on stellar evolution for SWPs. We
suspect that the stellar evolution history between SWPs and normal
stars could be somewhat different due to the presence of a planet, which
appears to affect only Li abundances, with no other elements
showing any difference.
As suggested by  \citet{King97}, the presence
of planets or associated circumstellar disks may affect a parent
star's initial angular momentum and/or subsequent evolution.
Specifically, the conservation of angular momentum in the protoplanetary disk
may induce increased mixing by causing rotational breaking in
the host stars during the pre--main--sequence evolution. This effect,
once it happens,  will destroy the Li completely. Considering
this suggestion as one possibility,
\cite{Israelian04} proposed a second hypothesis: that migration
triggers tidal forces and creates a shear instability that leads
to a strong depletion of Li abundance. Both mechanisms occur
during the pre--main--sequence evolution, and thus they can 
be used to explain
our result. Further theoretical and observational studies are very
desirable to investigate whether the two mechanisms are reasonable.

\subsection{Li Aabundances versus $\teff$}
Fig.~4 shows Li abundances versus temperatures for all stars. It
seems that Li abundance generally increases with increasing
$\teff$, and there is no significant gap between SWPs and normal
stars at a given temperature. Fig.~5 shows the stellar positions of
the two samples of stars in the HR diagram as compared with the
evolutionary tracks of \citet{VandenBerg00} with $\feh \sim +0.12$
dex. Stellar ages are also estimated from this comparison, and they
are presented in Table 1. The error of the ages due to the
uncertainties of $\teff$, $M_V$, and $\feh$ is about 15\%. It is
clear that our two samples of  stars are generally located in the same
regions of the HR diagram,  and during the selection all these stars are
classified as main--sequence stars with similar parameters.
However, the two stars from the comparison sample showing depleted Li
abundances, as indicated by downward--directed arrows in Fig.~5,
are actually subgiants. Other possible subgiant stars are not
evolved far away from the main sequence, and there is no
difference in Li abundance between SWPs and normal stars.

This observation that subgiants tend to have depleted Li
abundances reminds us to investigate the stellar positions
of SWPs at $5600 <\teff< 5850$ K in \citet{Israelian04}.
Here we redetermined the temperatures
for SWPs with $5600 <\teff< 5850$ K from \citet{Israelian04} in the
same way as \citet{Chen01}. In Fig.~3, the stellar
positions of these SWPs with new temperatures are plotted
against normal stars from \citet{Chen01} in this temperature range
with $\feh > -0.3$.
It is clear that a few SWPs in Israelian et al.'s (2004)
sample based
on the updated temperatures are
actually evolved subgiants, and downward--directed arrows in Fig.~3
indicate stars with undetected \ion{Li}{1} $\lambda$6708 lines.
The fact that evolved subgiants tend to have
depleted Li abundances is
consistent with the knowledge of stellar evolution theory.

It has been suggested in \citet{Ryan00} that the
Li abundances of subgiants can be somewhat depleted
due to the dilution proces without complete destruction,
as opposed to
main--sequence stars.
In this scenario, once a star evolves onto the
subgiant branch, it dilutes surface Li when convection brings
Li-poor material from deep layers to the surface.
This dilution is not obvious for slightly evolved subgiants
in the present study, and a few subgiants with 1.2 $\Msun$
show higher Li abundances than main--sequence stars.
Moreover, there is no difference
in Li abundance between planet hosts and normal stars on
the subgiant branch.
Actually, the dilution process mentioned above
is difficult to detect, and its effect is masked by
the large scatter of
Li abundances for even slightly evolved subgiants.
It is impossible to investigate how
the presence of a planet will affect lithium evolution for evolved
subgiants in the present work, since their Li abundances
have already been depleted due to their low
temperatures.

As shown in Fig.~4, there is a substantial scatter of Li
abundances at $\teff \sim 5800$ K. Considering the uncertainty
in temperature, it is difficult to investigate whether the scatter is
real. There seems to be a tendency that stars in the
upper envelope of the Li versus $\teff $ diagram seem to be slightly
younger than stars with the lowest Li abundances in the $\teff <5800$ K
range.  However, since there is a lack of young stars in the low--temperature
range and of old stars at the high--temperature edge in our sample, we
refrain from drawing any firm conclusions. A larger sample of stars
with different ages at a given temperature is needed to clarify
this issue.

\subsection{Li Anomalies of Stars With Planets and the Planet Formation}
The excess of Li depletion in SWPs in the
temperature range of $5600 <\teff< 5850$ K suggested
by \citet{Israelian04} is based on the fact that
Li abundances for SWPs are lower than for comparison stars.
A closer comparison shows that the
average of Li abundances for 24 SWPs at
this temperature range in \citet{Israelian04}
is $<0.99\pm0.45$, while this value is
 $2.08 \pm 0.49$ for the comparison stars from \citet{Chen01}.
In our work the average Li abundance is $<1.65 \pm 0.57$ for
SWPs versus $2.15 \pm 0.49$ for the comparison stars.
As presented above,
this anomaly in Li abundance of SWPs might reflect the effect of
planet presence on stellar evolution and cannot be used to
provide constraints on planet formation. In addition,
this effect is difficult to
detect from stars in the subgiant branch or later stages
if their Li abundances have already been depleted by other mechanisms.
Moreover, the
deviations of the average Li abundance between stars with and
without planets are 0.27 dex for $\teff > 5800$\, K
and 0.58 dex for $\teff < 5800$\,K. It seems that Li anomalies
concentrate in the range of $\teff < 5800$\,K, and thus they
do not relate to the Li--dip in the Hyades, which
happens in the temperature range of 6300--6600\,K as
first found by \citet{Boesgaard86}.

In connection with planet formation mechanism, our result does
not support the accretion scenario of planet formation.
Theoretically, \citet{Montalban02} predicted quantitative
estimates of the main--sequence evolution of stellar surface
lithium after planet ingestion. They found that the preservation
of $^7\mathrm Li$ occurs in a large mass range of 0.9--1.3 $\Msun$ at solar
metallicity. For SWPs with a typically super-solar metallicity,
the mass range is slightly reduced.
According to their theory, if a 1
$M_{\mathrm J}$ planet were dissolved in the convective envelope of an SWP
at a sufficiently old age for extra--mixing to be
inefficient, it would produce an enhancement of 0.3 dex in Li
abundance, which is unfortunately within the scatter of Li abundance
at the temperature range we investigated. But Li abundances for
stars with $\teff > 5900$ K in \citet{Israelian04} do not show
such a large scatter, and the similar abundance between SWPs and
the comparison stars argues against the accretion scenario.
Furthermore, most SWPs known so far have giant planets with masses larger than $5 M_{\mathrm J}$,
and if $10 M_{\mathrm J}$ were engulfed the enhancement would reach 0.9 dex, which
is easily detected.
Therefore, we favor the scenario that the surface
composition of SWPs is not polluted by the
accretion of planetary material, and the high mean metallicity
of SWPs is primordial.

\section{Conclusions}
\label{sect:conclusion}

We have made a parallel abundance analysis for 16 SWPs and 20
normal stars at the metallicity of $-0.2 < \feh < +0.4$ and the
temperature range of $5600 < \teff < 5900$ K. Our results
show that there is a higher probability for SWPs to have depleted Li abundances
than for comparison stars, which
confirms the possible Li abundance anomalies for SWPs at this temperature range
reported by \citet{Israelian04} after excluding evolved subgiants. We proposed that
the presence of planets around host stars may affect
stellar evolution by inducing additional mixing or shear instability,
which leads to a high probability of destroying surface lithium
while other elements are not affected. This effect can only
be detected for main--sequence SWPs since other depletion mechanisms
would not have started to act on main--sequence stars.
In view of this, the effect of planet presence should be considered
when the large scatters of the Galactic lithium in field disk stars
at the temperature range of $5600 < \teff < 5900$ K are investigated.

\section*{Acknowledgments}
This work is supported by the National Natural Science Foundation
of China under grants 10433010 and 10203002.


\begin{thebibliography}{}
\bibitem[Alonso et al. (1996)]{Alonso96} Alonso, A., Arribas, S. \&  Mart\'{\i}nez-Roger, C. 1996, A\&AS, 117, 227
{\bf 
\bibitem[Boesgaard \& Tripicco(1986)]{Boesgaard86} Boesgaard, A.M., \& Tripicco, M.J. 1986, ApJ, 302, L49
}
\bibitem[Carlsson et al.(1994)]{Carlsson94} Carlsson, M.,  Rutten, R.J., Bruls, J.H.M.J., \& Shchukina, N.G. 1994, A\&A, 288, 860
\bibitem[Cayrel (1988)]{Cayrel88} Cayrel, R. 1988, IAU symp., 132, 345
\bibitem[Chen et al.(2000)]{Chen00} Chen, Y.Q., Nissen, P.E., Zhao, G.
et al., 2000, A\&AS, 141, 491
\bibitem[Chen et al.(2001)]{Chen01} Chen, Y.Q., Nissen, P.E., Benoni, T., \& Zhao, G. 2001, \aap, 371, 943
\bibitem[Chen et al.(2002)]{Chen02} Chen, Y.Q., Nissen, P.E., Zhao, G., \& Asplund M. 2002, \aap, 390, 925
\bibitem[D'Antona \& Mazzitelli(1994)]{DAntona94} D'Antona, F., \& Mazzitelli, 1994, ApJS, 90, 467
\bibitem[Edvardsson et al.(1993)]{Edvardsson93} Edvardsson, B., Anderson, J., Gustafsson B., Lambert, D.L., Nissen, P.E., Tomkin, J., 1993, \aap, 275, 101
\bibitem[Gonzalez \& Laws(2000)]{Gonzalez00} Gonzalez, G., \& Laws, C. 2000, \apj, 119, 390
\bibitem[Gonzalez et al.(2001)]{Gonzalez01} Gonzalez, G., Lawsi, C., Tyagii, S., \& Reddy, B.E. 2001, AJ, 285, 403
\bibitem[Gustafsson et al.(1999)]{Gustafsson99} Gustafsson, B., Karlsson, T., Olsson, E., Edvardsson, B., \& Ryde, N. 1999, A\&A, 342, 426
\bibitem[Israelian et al.(2001)]{Israelian01} Israelian, G., Santos, N., Mayor, M, \& Rebolo, R. 2001, Nature, 411, 163
\bibitem[Israelian et al.(2004)]{Israelian04} Israelian, G., Santos, N., Mayor, M, \& Rebolo, R. 2004, \aap, 414, 211
\bibitem[King et al.(1997)]{King97} King, J.R., Deliysnnis, C.P., Hiltgen, D.D. et al., 1997, AJ, 113, 1871
\bibitem[Kurucz(1995)]{Kurucz95} Kurucz, R.L., 1995, CDROM, No. 23
\bibitem[Lambert et al.(1991)]{Lambert91} Lambert, D.L., Heath, J.E., \& Edvardsson, B., 1991, MNRAS, 253, 610
\bibitem[Mandell \& Ge(2004)]{Mandell04} Mandell, A.M., \& Ge, J. 2004, AJ, 127, 1147
\bibitem[Mayor \& Queloz(1995)]{Mayor95} Mayor, M., \&  Queloz, D. 1995, Nature, 378, 357
\bibitem[Montallb\'an \& Rebolo(2002)]{Montalban02} Montalb\'an, J., \& Rebolo, R. \aap, 386, 1043
\bibitem[Nordstr\"om et al.(2004)]{Nordstrom04} Nordstr\"om, B., Mayor, M., Andersen, J., et al., 2004, A\&A, 418, 989
\bibitem[Olsen(1983)]{Olsen83}
Olsen E.H., 1983, A\&AS 54, 55
\bibitem[Olsen(1993)]{Olsen93}
Olsen E.H., 1993, A\&A 102, 89
\bibitem[Perryman et al. (1997)]{Perryman97} Perryman, M.A.C. et al., 1997, The {\it Hipparcos} and Tycho Catalogs (ESA SP-1200, Noordwijk:ESA)
\bibitem[Prugniel \& Soubiran(2001)]{Prugniel01} Prugniel, P., \& Soubiran, C. 2001, A\&A, 369, 1048
\bibitem[Reddy et al.(2002)]{Reddy02} Reddy, B., Lambert, D.L., Laws, C., Gonzalez, G., \& Covery, K. 2002, \apj, 572, 1012
\bibitem[Ryan(2000)]{Ryan00} Ryan, R.G. 2000, \mnras, 316, L35
\bibitem[Santos et al.(2004)]{Santos04} Santos, N.C., Israelian, G., Garc\'ia L\'opex, G.J et al., 2004, \aap, 427, 1085
\bibitem[Schuster \& Nissen(1989)]{Schuster89} Schuster, B.W., \& Nissen, P.E. 1989, A\&A, 221, 65
\bibitem[Smith et al.(1998)]{Smith98} Smith, V.V., Lambert, D.L., \& Nissen, P.E. 1998, \apj, 506, 405
\bibitem[Takeda \& Kawanomoto (2005)]{Takeda05} Takeda, Y., \& Kawanomoto, S., 2005, PASJ, 57, 45
\bibitem[Valenti \& Fischer(2005)]{Valenti05} Valenti, J.A, \& Fischer, D.A., 2005, ApJS, 159, 141
\bibitem[VandenBerg et al.(2000)]{VandenBerg00} VandenBerg, D.A., Swenson, F.J., Rogers, F.J., Iglesias, C.A., \&  Alexander, D.R. 2000, \apj, 532, 430
\end{thebibliography}
\end{document}